\documentclass[prl,aps,twocolumn,showpacs,superscriptaddress,floatfix]{revtex4}

\usepackage{graphicx}
\usepackage{dcolumn}
\usepackage{bm}
\usepackage{color}

\RequirePackage[english]{babel}

\def\ds{\displaystyle}

\begin{document}

\bibliographystyle{apsrev}

\title{Conversion of terahertz wave polarization at the
boundary of a layered superconductor due to the resonance
excitation of oblique surface waves}

\author{Yu.O.~Averkov}
\affiliation{A.Ya.~Usikov Institute for Radiophysics and
Electronics, Ukrainian Academy of Sciences, 61085 Kharkov,
Ukraine}

\author{V.M.~Yakovenko}
\affiliation{A.Ya.~Usikov Institute for Radiophysics and
Electronics, Ukrainian Academy of Sciences, 61085 Kharkov,
Ukraine}

\author{V.A.~Yampol'skii}
\affiliation{A.Ya.~Usikov Institute for
Radiophysics and Electronics, Ukrainian Academy of Sciences, 61085
Kharkov, Ukraine} \affiliation{Advanced Science Institute, RIKEN, Saitama,
351-0198, Japan}

\author{Franco Nori}
\affiliation{Advanced Science Institute, RIKEN, Saitama, 351-0198, Japan}
\affiliation{Department of Physics, University of Michigan, Ann
Arbor, MI 48109, USA}

\begin{abstract}

We predict a complete TM$\leftrightarrow$TE transformation of the
polarization of terahertz electromagnetic waves reflected
from a strongly anisotropic boundary of a layered superconductor.
We consider the case when the wave is incident on the
superconductor from a dielectric prism separated from the sample
by a thin vacuum gap. The physical origin of the predicted phenomenon
is similar to the Wood anomalies known in optics, and is related to
the resonance excitation of the oblique surface waves. We also discuss the dispersion
relation for these waves, propagating along the boundary of the
superconductor at some angle with respect to the anisotropy axis,
as well as their excitation by the
attenuated-total-reflection method.

\end{abstract}





\maketitle

Surface waves play a very important role in many
fundamental resonant phenomena in different fields of modern physics.
A well-known example of such phenomena are the Wood
anomalies in the reflectivity and transmissivity of
periodically-corrugated metal and semiconducting samples (see, e.g.,
Refs.~\onlinecite{agr,rae,pet,bar}).
A more recent example is the extraordinary transmission of light through
metal films perforated by subwavelength holes~\cite{ebb,Marad}.
The excitation of surface plasmons can also result in an ``inverse'' effect of
resonant suppression of light transmission through perforated
metal films with thicknesses less than the skin-depth~\cite{katz}.

Recent interest on the effects listed above
is also partly due to their possible applications, e.g., in
photovoltaics, as well as the detection and filtering of radiation in
the far-infrared and visible frequency ranges. From this point
of view, it is interesting to consider layered
superconductors, instead of metals, for designing devices
which operate using the excitation of surface waves.
Indeed, the characteristic frequencies of surface waves in
these materials belong to the terahertz frequency range,
which is important for different applications, but is still
hard to reach for electronic and optical devices.

Layered superconductors are either artificially-grown stacks of
Josephson junctions, e.g., Nb/Al-AlO$_x$/Nb, or natural
high-temperature superconductors, such as $\rm
Bi_2Sr_2CaCu_2O_{8+\delta}$. These materials contain thin
superconducting layers separated by thicker dielectrics.
Experiments on the $\mathbf{c}$-axis transport in layered
superconductors justify the use of a theoretical model in which
the superconducting layers are coupled by the intrinsic Josephson
effect through the layers (see, e.g.,
Refs.~\onlinecite{Kl-Mu,Kl-Mu2}). Due to their layered structure, these
superconductors possess a strongly-anisotropic current capability.

The multi-layered structure of $\rm Bi_2Sr_2CaCu_2O_{8+\delta}$ (and
similar superconductors) supports the propagation of specific
Josephson plasma electromagnetic waves (JPWs) (see, e.g., the review
~\cite{Thz-rev} and references therein). The spectrum of JPWs lies above the
so-called Josephson plasma frequency, $\omega_J \sim 1$~THz. In a
semi-infinite sample, apart from bulk JPWs, surface Josephson
plasma waves (SJPWs) can propagate along the interface between
an external dielectric and a layered superconductor. These waves are
similar to the surface plasmon-polaritons in normal metals. As
shown in Refs.~\onlinecite{surf}, the
spectrum of SJPWs also lies in the terahertz range and consists
of two branches: one above $\omega_J$ and another below it. Similarly
to the \textit{optical} resonance phenomena in normal metals, the
resonance excitation of SJPWs is accompanied by Wood anomalies
in the reflection and transmission of \textit{terahertz waves}
through layered superconductors. However, the strong anisotropy of
the current capability can result in new resonance phenomena,
specific to layered superconductors.

In this paper, we predict one of such unusual resonance phenomena.
We show that the resonance excitation of surface waves can be accompanied
by a complete conversion of the polarization of the terahertz radiation
after reflection from an anisotropic boundary of a layered superconductor.
For example, if the incident wave has a transverse
magnetic (TM) polarization (a wave with the magnetic field parallel to
the sample surface), the reflected wave is completely transverse
electric (TE) (with the electric field parallel to the sample surface),
at an appropriate choice of the direction of the
incident wave vector. We emphasize that this phenomenon is not an analog
of the Brewster effect (where light with TM polarization, for a definite incidence angle, is perfectly transmitted through a
transparent dielectric surface, with no reflection, and, therefore, the
reflected light only contains the TE polarization). However,
in the phenomenon predicted here, the reflected wave has a TE
polarization when the incident radiation \textit{does not contain
the wave with this polarization}, i.e., when the incident wave is completely TM polarized. In other
words, the reflected wave is TE-polarized due to the \textit{conversion}
of the incident TM radiation, but not due to \textit{separation} of the TE wave
from the incident \textit{mixed} (TE+TM) radiation.

\begin{figure}
\includegraphics [width=8.0 cm,height=5.7 cm]{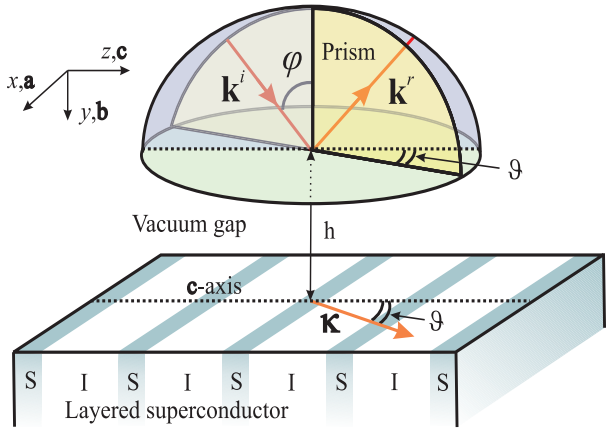}
\caption{\label{f1} (Color online) Schematic geometry of the problem for the excitation
of oblique surface waves. Here ${\bf k}^i$, ${\bf k}^r$, and
$\bm{\kappa}$ are the wave vectors of the incident, reflected, and oblique surface waves. Note that S=superconductor and I=insulator.}
\end{figure}

This polarization conversion has a clear physical explanation.
We study here the so-called ``attenuated-total-reflection'' (ATR) method for
producing surface waves. The surface wave is excited by the polarized
radiation (with, e.g., TM polarization) incident on the superconductor from
a dielectric semi-spherical prism separated from the sample by a thin vacuum
gap (see Fig. \ref{f1}).
We assume that the crystallographic \textbf{c}-axis of the superconductor lies
in the sample surface, and, therefore, the boundary of the superconductor is
strongly anisotropic. For arbitrary direction of the incident radiation, the
wave vector $\bm{\kappa}$ of the excited surface wave is oriented at some angle $\vartheta$
with respect to the anisotropy axis \textbf{c}, and we call these waves
oblique surface waves (OSWs)~\cite{av}. Due to the anisotropy, the OSWs contain all
components of the electric and magnetic fields and, therefore, the radiation reflected
from the superconductor contains waves of both, TM and TE, polarizations.
In other words, due to the anisotropy, the reflected wave is, generally, unpolarized.
Note that the ratio between the amplitudes
of the reflected TM and TE waves is controlled by the parameters of the problem,
such as the wave frequency $\omega$, angles $\vartheta$ and $\varphi$, the thickness $h$
of the vacuum gap, etc. The most important issue is the
possibility to choose the angles $\vartheta$ and $\varphi$ for the incident radiation
in such a way that the amplitude of the TM reflected wave vanishes. Recall that, in the isotropic
case, one can choose the optimal value of the gap thickness $h$ to provide the complete suppression
of the reflected wave under resonance conditions. Similarly, in the anisotropic case, we
can choose the optimal value of the angle $\vartheta$ to provide the complete suppression
of the reflected TM wave under resonance conditions. In this case, the desired complete conversion (TM $\rightarrow$ TE) of
the polarization takes place.

Below we discuss the dispersion relation for oblique surface waves, obtain
their excitations by the ATR method, and calculate the reflection coefficient $R_{\rm TM}$ for
the TM wave and the conversion coefficient $T_{{\rm TM} \rightarrow {\rm TE}}$ for
the mode conversion from TM to TE. Similar conversion phenomena can be
observed for transitions from TE to TM modes, from incident ordinary waves to
extraordinary ones, and from extraordinary waves to ordinary ones.

\begin{figure}
\includegraphics [width=8.0 cm,height=3.0 cm]{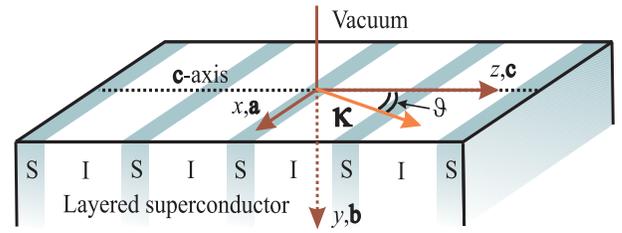}
\caption{\label{f2} (Color online) Geometry for oblique surface waves.}
\end{figure}

\textit{Oblique surface waves}.---
Consider a semi-infinite layered superconductor with the crystallographic \textbf{c}-axis  parallel to the sample surface (see Fig.~\ref{f2}).
We choose the coordinate system in such a way that the $z$-axis coincides with the crystallographic $\mathbf{c}$-axis of the superconductor and the $y$-axis is perpendicular to the sample surface. We study oblique surface waves, propagating along the interface vacuum-layered superconductor at an angle $\vartheta$ with respect to the \textbf{c}-axis, with electric $\mathbf{E}$ and magnetic $\mathbf{H}$ fields in the form, $$\mathbf{E}(x,y,z,t),\:\mathbf{H}(x,y,z,t)\propto \exp[i(\bm{\kappa}\bm{\rho} -\omega t)],$$
where $\bm{\rho}=(x,z)$ is the radius vector in the $xz$-plane,
$\bm{\kappa}=(k_x, k_z)$ is the wave vector of the oblique surface wave, $|\bm{\kappa}|>\omega/c$, and $c$ is the speed of light.
It is suitable to present the electromagnetic field as a sum of two terms that correspond to the ordinary and extraordinary evanescent waves,
\begin{equation}\label{e2}
\mathbf{E}(y)=\mathbf{E}^{\rm (o)}(y)+\mathbf{E}^{\rm (e)}(y), \ \:\mathbf{H}(y)=\mathbf{H}^{\rm (o)}(y)+\mathbf{H}^{\rm (e)}(y).
\end{equation}
The electric field of the ordinary wave and the magnetic field of the extraordinary wave are orthogonal to the anisotropy $z$-axis:
\begin{equation}\label{pol}
\mathbf{E}^{\rm (o)}_z(y)=\mathbf{H}^{\rm (e)}_z(y)=0.
\end{equation}
For each of these waves, the Maxwell equations give the relations between the components of the electric $\mathbf{E}^{\rm vac}$ and magnetic $\mathbf{H}^{\rm vac}$ fields in the vacuum and the same exponential law for the decay of both waves along the $y$-axis,
\begin{equation}\label{e3}
\mathbf{E}_{\rm vac}^{\rm (o)}(y), \;\mathbf{E}_{\rm vac}^{\rm (e)}(y), \;\mathbf{H}_{\rm vac}^{\rm (o)}(y), \;\mathbf{H}_{\rm vac}^{\rm (e)}(y) \propto \exp(-ik_{y}^{\rm vac} y),
\end{equation}
with imaginary normal component of the wave vector,
\begin{equation}\label{k-vac}
k_{y}^{\rm vac}=i(k_x^2 + k_z^2 - \omega^2/c^2)^{1/2}.
\end{equation}

The electromagnetic field inside the layered superconductor
is determined by the distribution of the gauge-invariant phase difference $\chi$ of the order parameter
between neighboring layers. This phase difference can be
described by a set of coupled sine-Gordon equations (see review~\cite{Thz-rev} and references therein). In the continuum and linear approximation, $\chi$  can be excluded from the set of equations for
electromagnetic fields, and the electrodynamics of layered superconductors can be described in terms of an anisotropic
frequency-dependent dielectric permittivity with
components $\varepsilon_c$ and $\varepsilon_{ab}$ across and along the layers,
respectively~\cite{negref},
\begin{equation}\label{epsilon2}
\begin{array}{l}
\varepsilon_{c}(\Omega)=\varepsilon\left(1-\ds\frac{1}{\Omega^2}+i\nu_{c}
\ds\frac{1}{\Omega}\right),\vspace{0.2cm} \\
\varepsilon_{ab}(\Omega)=\varepsilon\left(1-\ds\frac{1}{\Omega^2}
\gamma^2+i\nu_{ab}\ds\frac{1}{\Omega}\gamma^2\right).
 \end{array}
\end{equation}
Here we introduce the dimensionless parameters
$\Omega=\omega/\omega_J$, $\nu_{ab}= 4\pi
\sigma_{ab}/\varepsilon\omega_J\gamma^2$, and $\nu_{c}= 4\pi
\sigma_{c}/\varepsilon\omega_J$, $\gamma=\lambda_c/\lambda_{ab} \gg 1$ is the
current-anisotropy parameter, $\lambda_c
=c/\omega_J\varepsilon^{1/2}$ and $\lambda_{ab}$ are the
magnetic-field penetration depths along and across the layers,
respectively. The relaxation frequencies
$\nu_{ab}$ and $\nu_{c}$ are proportional to the averaged
quasi-particle conductivities $\sigma_{ab}$ (along the layers) and
$\sigma_{c}$ (across the layers), respectively; $\omega_J = (8\pi
e D j_c/\hbar\varepsilon)^{1/2}$ is the Josephson plasma
frequency. The latter is determined by the critical Josephson
current density $j_c$,  the interlayer dielectric constant
$\varepsilon$, and the spatial period $D$ of the layered structure.

Contrary to the vacuum, the Maxwell equations give different laws for the decays of the ordinary and extraordinary waves in layered superconductors,
\[
\mathbf{E}_{\rm s}^{\rm (o)}(y), \;\mathbf{H}_{\rm s}^{\rm (o)}(y) \propto \exp(ik_{{\rm s}y}^{\rm (o)}y)\:,
\]
\begin{equation}\label{e4}
\mathbf{E}_{\rm s}^{\rm (e)}(y), \;\mathbf{H}_{\rm s}^{\rm (e)}(y) \propto \exp(ik_{{\rm s}y}^{\rm (e)}y)
\end{equation}
with \begin{equation}\label{HTS5}
k_{{\rm s}y}^{\rm (o)}=i(k_x^2+k_z^2-\omega^2\varepsilon_{ab}/c^2)^{1/2}\:,
\end{equation}
\begin{equation}\label{HTS5a}
k_{{\rm s}y}^{\rm (e)}=i(k_x^2+k_z^2 \varepsilon_c/\varepsilon_{ab}-\omega^2\varepsilon_{c}/c^2)^{1/2}\:.
\end{equation}

\begin{figure}
\includegraphics [width=8 cm,height=5.75 cm]{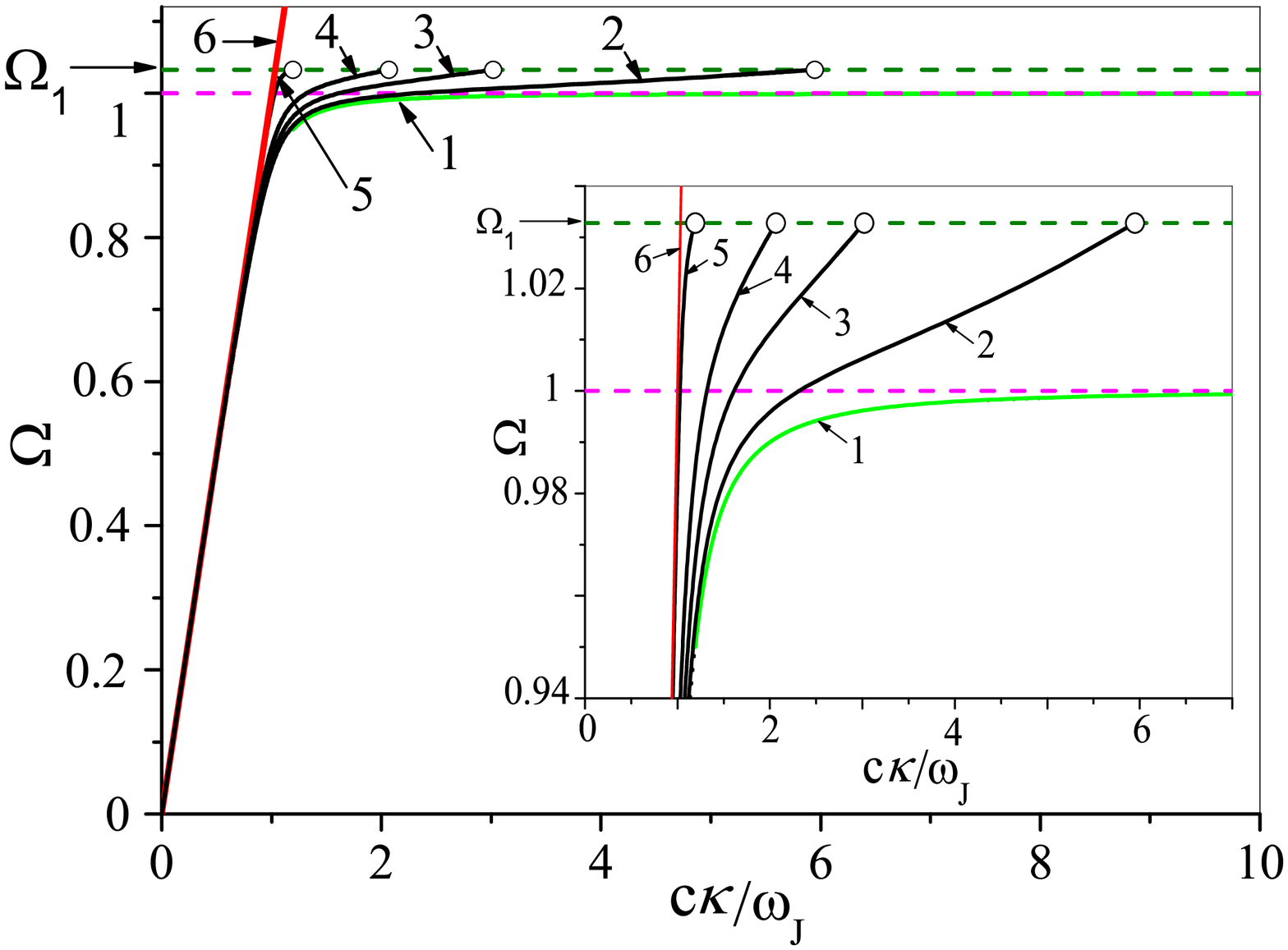}
\caption{\label{f3} (Color online) Dispersion curves $\Omega(c\kappa/\omega_J)$ of the OSWs for $\varepsilon=16$, $\gamma=200$, and $\nu_{ab}=\nu_c=0$. Curves 1--6 correspond to $\vartheta=0^{\circ}$, $10^{\circ}$, $20^{\circ}$, $30^{\circ}$, $60^{\circ}$, and $90^{\circ}$, respectively. The hollow circles at $\Omega=\Omega_1=\sqrt{\varepsilon/(\varepsilon-1)}$ show the end-points of the dispersion curves. In these points, the extraordinary wave, which is part of the OSW, is transformed into a bulk wave.}
\end{figure}

From the continuity conditions for the tangential
components of the electric and magnetic fields at the interface $y=0$,
we derive the dispersion equation of the OSWs,
\begin{eqnarray}
 &   k_x^2 k_z^2 (\varepsilon_{ab}-1)\displaystyle\frac{k_{{\rm s}y}^{\rm (o)}+k_{y}^{\rm vac}}{k_{{\rm s}y}^{\rm (o)} k_{y}^{\rm vac}}\nonumber \\
&    -\Bigl[k_z^2(\varepsilon_{ab}-1)+\varepsilon_{ab}k_{y}^{\rm vac} (k_{y}^{\rm vac}+k_{{\rm s}y}^{\rm (e)})\Bigr]\nonumber \\
&    \times\Bigl[\displaystyle\frac{1}{k_{{\rm s}y}^{\rm (o)}}\Bigl(\frac{\omega^2}{c^2}\varepsilon_{ab}-k_z^2\Bigr)+
    \displaystyle\frac{1}{k_{y}^{\rm vac}}\Bigl(\frac{\omega^2}{c^2}-k_z^2\Bigr)\Bigr]=0.
\end{eqnarray}
Figure~\ref{f3} illustrates the numerically calculated dispersion curves of the OSWs for
different values of the angle $\vartheta$.

\textit{Excitation of the OSWs and the polarization conversion}.--- Here we investigate the OSWs excitation using the ATR method.
Consider a TM polarized wave incident from a dielectric prism with permittivity $\varepsilon_p$
onto a layered superconductor separated from the prism by a vacuum gap of thickness $h$ (see Fig. \ref{f1}). The
incident angle $\varphi$ exceeds the limit angle $\theta_t =
\arcsin(\varepsilon_p^{-1/2})$ for the total internal reflection. In this case the incident wave can
excite a surface wave if the resonance
condition, $\omega \varepsilon_p^{1/2} \sin\varphi=c\kappa$ is satisfied.

The electromagnetic field in the dielectric prism is a sum of three terms that correspond to the incident TM polarized wave and two reflected waves with the TM and TE polarizations. Thus, the $x$-component of the electric field in the prism can be presented as
\[
E_x^p (x,y,z,t) = \left[E^i_x\exp(ik_{p\,y} y) \right.
\]
\begin{equation}\label{eq1}
\left.+ (E^r_{x \:{\rm TM}}+E^r_{x \:{\rm TE}})\exp(-ik_{p\,y} y) \right]\exp\bigl[i(\bm{\kappa}\bm{\rho} -\omega t)\bigr]
\end{equation}
where $E^i_x$, $E^r_{x \:{\rm TM}}$, and $E^r_{x \:{\rm TE}}$ are the amplitudes of the $x$-components of the electric field for the incident waves and for the reflected TM and TE waves, respectively; $k_{p\,y} =(\omega /c)\,\varepsilon_p^{1/2} \cos\varphi$ is the normal component of the wave vector of the incident and reflected waves. Other components of the electromagnetic field in the prism are expressed via the amplitudes $E^i_x$, $E^r_{x \:{\rm TM}}$, and $E^r_{x \:{\rm TE}}$ using the Maxwell equations.

\begin{figure}
\includegraphics [width=8.0 cm,height=5.75 cm]{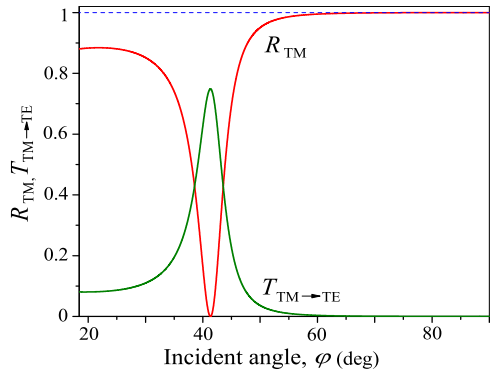}
\caption{\label{f4} (Color online) The reflectivity $R_{\rm TM}$ and the transformation coefficient $T_{{\rm TM}\rightarrow {\rm TE}}$, both versus $\varphi$.}
\end{figure}

\begin{figure}
\includegraphics [width=8.0 cm,height=5.75 cm]{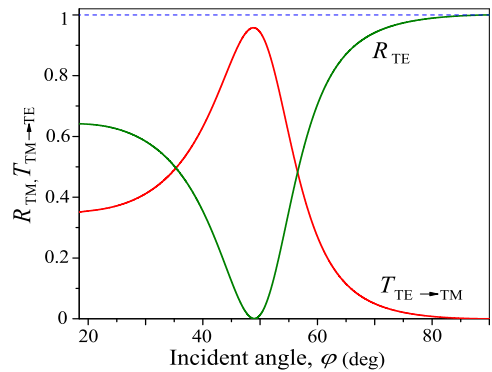}
\caption{\label{f5} (Color online) The reflectivity $R_{\rm TE}$ and the transformation coefficient $T_{{\rm TE}\rightarrow {\rm TM}}$, both versus $\varphi$.}
\end{figure}

In the vacuum gap, we define the electromagnetic field as the
superposition of four evanescent waves. They are two TM modes with imaginary $y$-components $\pm k_{y}^{\rm vac}$ of the wave vectors given by Eq.~(\ref{k-vac}) and two TE modes with the same wave vectors. In the layered superconductor we have, similarly to the previous section, ordinary and extraordinary waves with the normal components of the wave vectors given by Eqs.~(\ref{HTS5}) and (\ref{HTS5a}). Then, using the continuity boundary conditions for the tangential components of the electric and magnetic fields at the interfaces dielectric-vacuum and vacuum-layered superconductor, we obtain eight linear algebraic equations for eight unknown amplitudes (for 4 waves in the vacuum, 2 waves in the layered superconductor, and amplitudes $E^r_{x \:{\rm TM}}$, $E^r_{x \:{\rm TE}}$ of the reflected waves in the dielectric prism). Solving these equations, one can derive the reflectivity coefficient $R_{\rm TM}(\varphi, \vartheta)=|\textbf{E}^r_{\:{\rm TM}}|^2/|\textbf{E}^i|^2 = |E^r_{x \:{\rm TM}}/E^i_x|^2$ for the TM wave and the coefficient $T_{{\rm TM}\rightarrow {\rm TE}}(\varphi, \vartheta)=|\textbf{E}^r_{\:{\rm TE}}|^2/|\textbf{E}^i|^2 = |E^r_{x \:{\rm TE}}/E^i_x|^2 \tan^2\vartheta \cos^2\varphi$ of the transformation from the TM mode to the TE one. Figure~\ref{f4} shows the dependences of these coefficients on the angle $\varphi$, calculated for $\varepsilon_p=10$, $\varepsilon=16$, $\gamma=200$, $\Omega=1.00035$, $\nu_{ab}=\nu_c=10^{-3}$, $h=c/2\omega_J$, $\vartheta=15.6^{\circ}$.
The curves demonstrate the complete transformation of the TM-polarized wave to the TE wave, for the optimal value of the angle $\vartheta=15.6^{\circ}$, under resonance condition. The positions of the minimum in the $R_{\rm TM}(\varphi)$ dependence and of the maximum in the $T_{{\rm TM}\rightarrow {\rm TE}}(\varphi)$ dependence coincide. This position corresponds to the most effective excitation of the oblique surface wave.
Similar effects can be also observed for the TE$\rightarrow$TM complete transformation of the polarization, as well as for the complete transformation of the incident ordinary waves to extraordinary and vice versa. Figure~\ref{f5} shows the dependences of the reflection coefficient $R_{\rm TE}(\varphi, \vartheta)$ and the coefficient $T_{{\rm TE}\rightarrow {\rm TM}}(\varphi, \vartheta)$ of the transformation from the TE mode to the TM one calculated for $\varepsilon_p=10$, $\varepsilon=16$, $\gamma=200$, $\Omega=1.0004$, $\nu_{ab}=\nu_c=10^{-3}$, $h=0.18 c/\omega_J$, and $\vartheta=22^{\circ}$.

\textit{Conclusions}.--- We predict the complete polarization conversion for terahertz waves reflecting from the strongly-anisotropic surface of layered superconductors. The origin of this unusual effect is similar to the Wood anomalies known in optics, and is related to
the resonance excitation of the oblique surface waves.

We acknowledge partial support from the NSA, LPS, ARO, NSF grant
No.~0726909, JSPSRFBR Contract No.~09-02-92114, Kakenhi (S), MEXT,
the JSPS-FIRST program, Ukrainian State Program on Nanotechnology, and the Program FPNNN of NAS of Ukraine (grant No~9/11-H).

\end{document}